\documentclass[structabstract]{aa}  
\usepackage{aalongtable}
\usepackage{natbib}
\bibpunct{(}{)}{;}{a}{}{,} 
\usepackage{graphicx}
\usepackage{txfonts}
\begin{document}
   \title{Non-thermal radio emission from O-type stars}
   \subtitle{V. 9~Sgr}

   \author{R. Blomme
           \and D. Volpi
          }


   \institute{Royal Observatory of Belgium,
              Ringlaan 3, B-1180 Brussel, Belgium \\
              \email{Ronny.Blomme@oma.be}
            }

   \date{Received date; accepted date}

 
  \abstract
   {The colliding winds in a massive binary system generate synchrotron
emission due to a fraction of electrons that have been accelerated to
relativistic speeds around the shocks in the colliding-wind region.
   }
   {We studied the radio light curve of 9~Sgr = HD 164794, a massive O-type
binary with a 9.1-yr period. We investigated whether the radio emission
varies consistently with orbital phase and we determined some parameters
of the colliding-wind region.
   }
   {We reduced a large set of archive data from the Very Large Array (VLA)
to determine the radio light curve of 9~Sgr at 2, 3.6, 6 and 20~cm. We also
constructed a simple model that solves the radiative transfer in the
colliding-wind region and both stellar winds.
   }
   {The 2-cm radio flux shows clear phase-locked variability with the orbit.
The behaviour at other wavelengths is less clear, mainly due to a lack
of observations centred on 9~Sgr
around periastron passage. The high fluxes and nearly flat
spectral shape of the radio emission show that synchrotron radiation
dominates the radio light curve at all orbital phases. The model provides
a good fit to the 2-cm observations, allowing us to estimate that the brightness
temperature of the synchrotron radiation emitted in the colliding-wind
region at 2 cm is at least $4\times 10^8$~K.
   }
   {The simple model used here already allows us to derive important information
about the colliding-wind region. We propose that 9~Sgr is a good candidate
for more detailed modelling, as the colliding-wind region remains adiabatic
during the whole orbit thus simplifying the hydrodynamics.
   }


   \maketitle
%

\section{Introduction}

In a massive-star binary the stellar winds collide, forming two shocks,
one on each side of the contact discontinuity. In this colliding-wind region 
(CWR) the material gets heated and compressed \citep{1993ApJ...402..271E}. 
If the binary has a sufficiently long period, the time scale for radiative
cooling of the material will far exceed the flow time scale,
and the CWR will not 
be able to cool down. The CWR is thus in the adiabatic regime. The
high temperature leads to an excess
of X-ray emission, which also has a harder spectrum than the 
X-ray contribution of the
stellar winds themselves \citep{1992ApJ...386..265S, 2012ASPC..465..301G}. 

The CWR also has an important effect on
the radio continuum emission. Around the shocks, the Fermi mechanism 
accelerates a fraction of the electrons up to relativistic speeds
\citep{1993ApJ...402..271E}. These electrons then emit
synchrotron radiation as they spiral in the magnetic field. 
This synchrotron radiation can be detected at radio wavelengths. 
It can be recognized 
by an excess flux compared to that expected from the free-free wind
emission, 
by variability that is locked to the orbital phase, 
by a high brightness temperature,
and by a non-thermal spectral
index\footnote{The spectral index is the quantity
               $\alpha$ in $F_\nu \propto \nu^\alpha$.}
\citep{1989ApJ...340..518B}.
\citeauthor{1989ApJ...340..518B} 
consider a spectral index $\alpha < 0.0$ to be a clear
indicator of non-thermal emission, as it is significantly different
from the $\alpha=+0.6$ expected for thermal emission in
a spherically symmetric stellar wind.

Previous papers in this series have studied a number of these non-thermal
radio emitters. 
For \object{HD 168112}, we found that the radio fluxes show periodic behaviour,
suggesting that it is a binary system, though this still awaits confirmation
by spectroscopic observations \citep{2005A&A...436.1033B}.
For \object{Cyg~OB2\,\#9}, we also detected periodic behaviour in the radio
fluxes \citep{2008A&A...483..585V}, 
while \citet{2008A&A...483..543N} used spectroscopic observations to
show the binary nature of this system. Later, we monitored the 
periastron passage of Cyg~OB2\,\#9 in detail, allowing us to model the
system \citep{2012A&A...546A..37N, 2013A&A...550A..90B}.
For \object{Cyg~OB2\,\#8A} (a known binary) 
we showed that the radio data are better
explained by assuming the secondary component to have the stronger wind
\citep{2010A&A...519A.111B}.
The radio variability of \object{HD 167971} (a triple system) suggests that we
are detecting the CWR between the binary at the core of the system and the
third component further away \citep{2007A&A...464..701B}.

In this paper, we study 9~Sgr = \object{HD 164794}. Proof that 9~Sgr is a binary
was a long time coming. This allowed for the possibility that the
non-thermal radio emission was somehow intrinsic to the stellar wind
of 9~Sgr itself. In a single star, the shocks due to the intrinsic
instability of the radiation driving mechanism would be the sites where
the Fermi acceleration mechanism operates \citep{1985ApJ...289..698W}.
However, \citet{2006A&A...452.1011V} showed that single-star winds
are unlikely to explain the non-thermal radio emission. 

First hints of the binary nature of 9~Sgr came from the excess of
X-ray emission and the long-term radial velocity variations in the 
optical spectra \citep{2002A&A...394..993R}. Later,
\citet{2005mshe.work...85R} showed 9~Sgr to have a clear SB2 signature,
but the period could only be estimated.
Finally, \citet{2012A&A...542A..95R} 
showed that 9~Sgr is a spectroscopic binary,
with an O3.5~V((f$^+$)) primary and an O5--5.5~V((f)) secondary.
It has a highly eccentric orbit ($e=0.7$) with a period that 
\citeauthor{2012A&A...542A..95R} determined to be 8.6 years,
but which has recently been revised to 9.1 year (Rauw, 2013,
pers. comm.). The long period of this system explains
why it was so difficult to prove the binarity.

Among the O+O binaries, 9~Sgr is the system
with the longest known period. A study of its colliding-wind region
is therefore important as it samples a significantly different part
of the parameter space of colliding-wind binaries.

\citet{1980ApJ...238..196A} were the first to detect 9~Sgr at radio
wavelengths. Their flux value, interpreted in a strictly thermal wind
model, led to a radio mass-loss rate a factor 40 higher than that
from H$\alpha$ and the ultraviolet P Cygni profiles.
A second observation showed some possible variability 
\citep{1981ApJ...250..645A}. Further monitoring allowed
\citet{1984ApJ...280..671A} to conclude that 9~Sgr is a non-thermal
radio emitter, because of the non-thermal spectral index and
the flux variability. This was confirmed by additional data
by \citet{1989ApJ...340..518B}.
A clear non-thermal spectral index was also found in the radio data
analysed by \citet{2002A&A...394..993R}.
They modelled the radio observations under the then current assumption
that the star was single.

In this paper, we use data from the Very Large Array (VLA) archive to study 
the radio light curve of 9~Sgr.
In Sect.~\ref{section data reduction} we present the data and their
reduction. We discuss the resulting radio light curve in 
Sect.~\ref{section radio light curve}.
In Sect.~\ref{section modelling} we present the model for the colliding-wind
region which we used to interpret the observations. We discuss the 
results of this modelling in
Sect.~\ref{section discussion}, and we present our conclusions in
Sect.~\ref{section conclusions}.

\section{Data reduction}
\label{section data reduction}

We selected all data from the VLA archive that were centred on, or close to,
9~Sgr.
The data found cover a range of 24 years.
We reduced the visibility data using the Astronomical Image
Processing System (AIPS), developed by the NRAO. We applied
the standard procedures for antenna gain calibration, absolute flux
calibration, imaging and deconvolution. 
The absolute flux calibration uses a model for the flux calibrator when
that is available.
For details of the data reduction, we refer
to the previous papers in this series 
\citep{2005A&A...436.1033B, 2007A&A...464..701B, 2008A&A...483..585V,
2010A&A...519A.111B}.

When the VLA is in one of the configurations with lower spatial resolution,
the resulting images are dominated by the presence of 
the \object{Hourglass Nebula}. 
This is a blister-type H~{\sc ii} region ionized by the 
O7.5~V star \object{Herschel 36} \citep{2010MNRAS.407.1170K}. 
It is a much stronger radio source than 9~Sgr itself.
At 6~cm, in the configuration with the lowest spatial resolution, 
the flux of the Hourglass Nebula is 0.7 -- 1.7~Jy (depending on the exact
configuration). At 20~cm, the flux is 3.6 -- 4.3~Jy. This is to
be compared to the 9~Sgr flux which is of order 1 -- 10 mJy.
The problem with the high Hourglass Nebula
flux does not occur at higher spatial resolutions because of
the filtering properties of a radio interferometer: the 6-cm
flux is only $\sim$~5 mJy and the 20 cm one 20 -- 30 mJy for the 
configuration with the highest spatial resolution.

The high flux of the Hourglass Nebula considerably complicates the 
detection of 9~Sgr and the measurement of its flux. 
We therefore modelled this source separately and then
subtracted it directly from the visibility data.
Further background is removed by systematically dropping visibility data
on the shortest baselines.
In this way, we typically achieve a 1-sigma 
noise level of 2 -- 10~mJy at 
6~cm and 30 -- 50 mJy at 20~cm for the configurations
with the lowest spatial resolution, but these values are still
too high to allow a detection. 
It is only at the higher
spatial resolution configurations that we can detect 9~Sgr.
For those data sets where 9~Sgr is detected, we further applied
a single round of phase-only self-calibration.

We measured the fluxes and error bars by fitting an
elliptical Gaussian to the source on the images. The results
are listed in the Appendix in Table~\ref{table radio data}.
The error bars 
include not only the rms noise
in the map, but also an estimate of the systematic errors that 
were evaluated using a jack-knife technique. This technique drops
part of the observed data and re-determines the fluxes, giving some
indication of systematic errors that are present.
Note that the absolute calibration uncertainty is not included in the
error bars listed in the table. 
These uncertainties are estimated at 1--2 \% at 20, 6, and 3.6~cm, 
and 3--5 \% at 2 and 0.7~cm \citep{VLA_Calibrator_Manual}.
Where the source could not be detected, we assigned an upper limit
of three
times the rms noise around the measured position.

A number of observations are not centred on 9~Sgr, but on another,
nearby, target.
The 9~Sgr flux values and error bars for these observations
have been corrected for the decreasing sensitivity of the primary beam
and for the increased size a point source has due to bandwidth
smearing \citep{1999ASPC..180..371B}. As the bandwidth smearing
effect spreads out the flux over a larger area, the flux measurement
becomes more difficult and the values should therefore be
considered as less reliable.

When there are multiple targets for one observation, we list
in Table~\ref{table radio data} only
that one with the best detection or the lowest noise level (in case
of a non-detection). This is usually
the one where the
field centre is closest to 9~Sgr.
The exceptions are the two AF399 3.6-cm observations, where the integration
time on the offset position is much longer than on 9~Sgr, resulting
in a better image.
We further exclude from Table~\ref{table radio data}
those observations with upper limits higher than 50 mJy (which
corresponds to about 4 times the highest detection at all wavelengths).

To further increase the signal, we also combined data that were
taken close-by in time. This procedure of course assumes that
there are no significant changes on short timescales.
As we are looking for variability related to the 9.1-year
orbital period, we combined data that were taken less than
1 month apart.

\begin{table}
\caption{Comparison of our radio fluxes with those in the literature.}
\label{table comparison literature}
\begin{tabular}{lllccc}
\hline
\hline
& \multicolumn{1}{c}{ID} & \multicolumn{1}{c}{date} & \multicolumn{1}{c}{literature flux} & \multicolumn{1}{c}{our flux} & \multicolumn{1}{c}{ref} \\
&    &      & \multicolumn{1}{c}{(mJy)} & \multicolumn{1}{c}{(mJy)} & \\
\hline
\multicolumn{3}{l}{{\bf 2 cm}} \\
& BIEG   & 1982-02-09 & $<$2.4        &  $<$2            & 1 \\
& AC42   & 1983-08-22 & $<$0.8        &  $<$0.9          & 1 \\
& AC116  & 1985-02-16 & 0.7 $\pm$ 0.1 &  1.01 $\pm$ 0.28 & 1 \\
\multicolumn{3}{l}{{\bf 3.6 cm}} \\
& AB1005 & 2001-03-08 & 1.6 $\pm$ 0.4 &  3.6  $\pm$ 0.9  & 2 \\
\multicolumn{3}{l}{{\bf 6 cm}} \\
& CHUR   & 1979-07-13 & 1.0 $\pm$ 0.4 &  9.0  $\pm$ 2.7  & 1 \\
& CHUR   & 1980-05-22 & 1.8 $\pm$ 0.3 &  1.82 $\pm$ 0.47 & 1 \\
& BIEG   & 1982-02-09 & 2.5 $\pm$ 0.3 &  1.70 $\pm$ 0.28 & 1 \\
& BIGN   & 1982-05-26 & 2.4 $\pm$ 0.3 &  2.76 $\pm$ 0.11 & 1 \\
& AC42   & 1983-08-22 & 1.5 $\pm$ 0.2 &  2.20 $\pm$ 0.14 & 1 \\
& AC116  & 1984-11-27 & 2.0 $\pm$ 0.2 &  1.94 $\pm$ 0.22 & 1 \\
& AB327  & 1985-01-28 & 1.5 $\pm$ 0.1 &  2.08 $\pm$ 0.19 & 1 \\
& AC116  & 1985-02-16 & 1.9 $\pm$ 0.1 &  1.96 $\pm$ 0.13 & 1 \\
& AB1005 & 2001-03-08 & 2.8 $\pm$ 0.4 &  2.9  $\pm$ 0.74 & 2 \\
\multicolumn{3}{l}{{\bf 20 cm}} \\
& BIGN   & 1982-05-26 & 3.6 $\pm$ 0.3 &  5.23 $\pm$ 0.20 & 1 \\
& AC42   & 1983-08-22 & 3.9 $\pm$ 0.4 &  3.24 $\pm$ 0.37 & 1 \\
& AB1005 & 2001-03-08 & 4.5 $\pm$ 1.2 & 12.7  $\pm$ 3.6  & 2 \\
\hline
\end{tabular}
\tablebib{(1)~\citet{1989ApJ...340..518B}; (2)~\citet{2002A&A...394..993R}.}
\end{table}

Table~\ref{table comparison literature}
shows a comparison with 
values previously published in the literature.
Our current values of the AB1005 3.6 and 20-cm fluxes are very different
from those of our previous reduction of the same data given in
\citet{2002A&A...394..993R}.
In that paper, we used the less sophisticated technique
of removing data on short and intermediate-length baselines
to remove the effect of the Hourglass nebula.
Our current technique of subtracting the Hourglass nebula visibilities
from the original data is better adapted to handle the
problem of this strong, nearby source.
The flux values presented here
therefore supersede those published in \citeauthor{2002A&A...394..993R}

\citet{1989ApJ...340..518B} list a number of flux determinations
based on data in common with those discussed here. The major difference
found is in the CHUR (1979-07-13) 6~cm flux.
\citeauthor{1989ApJ...340..518B}  give a value of
$1.0 \pm 0.4$~mJy, while we find $9.0 \pm 2.7$~mJy. Details of their data
reduction are given in \citet{1980ApJ...238..196A}, who 
point out the problems
in the data reduction due to the presence of \object{M~8} (or, more correctly,
the Hourglass nebula). Apparently, the procedure they
used to remove the effect of the nebula is not based on subtracting the
Hourglass nebula visibilities. We therefore have greater confidence in the
results presented here. Other differences between our results and 
those of \citeauthor{1989ApJ...340..518B} are smaller, or not significant.
We again attribute any significant differences to the different techniques
to remove the effect of the Hourglass nebula.

\section{Radio light curve}
\label{section radio light curve}

\begin{figure}
\resizebox{\hsize}{!}{\includegraphics{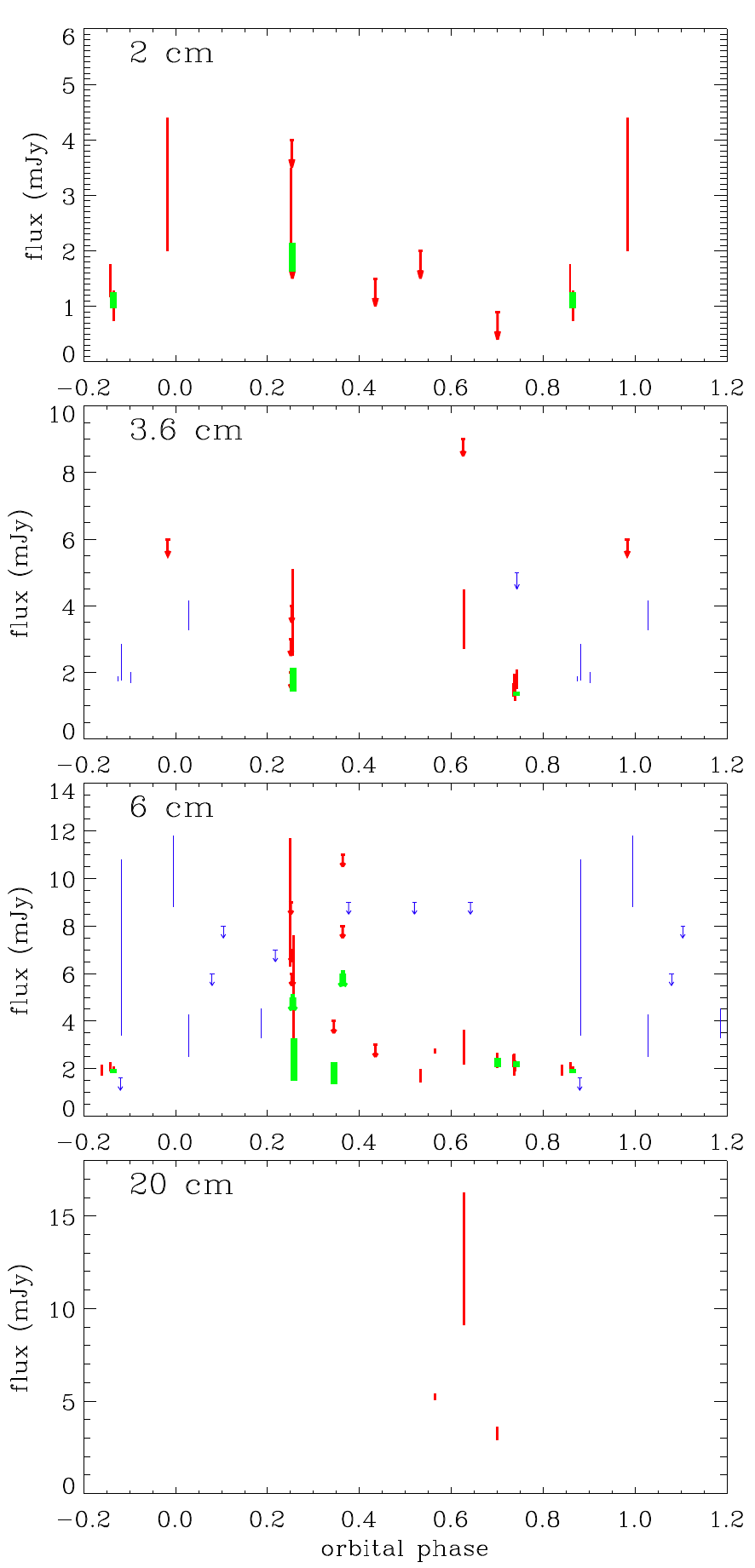}}
\caption{9~Sgr radio fluxes as a function of orbital phase in the
9.1-year period. The detections are shown with their error bars, the
upper limits as arrows. Red data points are observations that 
are centred on 9~Sgr, blue data points have 9~Sgr off-centre. 
The green bars indicate
data that have been combined. Phase 0.0 corresponds to periastron passage.}
\label{fig fluxes}
\end{figure}

We first tried to determine a period from our radio data, using the
string-length method \citep{1983MNRAS.203..917D}. We normalized the
fluxes to their maximum value, 
so that the flux differences and phase differences
have an equal weight. We tried the string-length method with
various possibilities: using data at
all wavelengths, or only the 6-cm data (as most of our data are at 6 cm),
including the observations where 9~Sgr was off-centre, or not. No combination
however leads to a period near that of the spectroscopically determined
one. As the quality of the spectroscopic data is much better than the
radio data we have, from hereon we use the 9.1-year period as determined
by Rauw (2013, pers. comm.).
Note that this value for the period updates the one given by
\citet{2012A&A...542A..95R}.

Figure~\ref{fig fluxes} plots the observed radio fluxes at 2, 3.6, 6 and
20~cm, as a function of phase in the 9.1-year orbit. The 2-cm fluxes
clearly show variability: around periastron ($\pm 0.2$ in phase) the fluxes
are high (around 3 mJy), but in the phases in between the fluxes are only
1--2 mJy. For 6~cm, observations centred on 9~Sgr at periastron
are lacking. Based on the off-centre data (which are less
reliable), there 
is a suggestion that the fluxes
around periastron are high. Away from periastron the
fluxes are consistently low (about 2 mJy).
For 3.6 cm the picture is less clear: there is also variability,
but there is no clear indication of a flux increase near periastron.
For 20~cm, the only high flux value is quite some distance away from
periastron, and it is surrounded on each side by a lower-flux
observation. Other wavelengths (0.7 and 90~cm) have only a single observation 
with a high upper limit.

The 2-cm behaviour is qualitatively in agreement with what is expected
for an eccentric long-period binary. At all orbital phases the stellar
winds collide, leading to the generation of synchrotron emission. 
As the system approaches periastron, the stars move
much closer together, leading to a more energetic collision
and therefore more
synchrotron emission. In principle, this synchrotron emission could
be partially or completely absorbed by the free-free absorption of the
material in the stellar winds. However, 
because of the long period of this system,
the stars will be far apart and we would therefore expect little
effect of the stellar wind absorption.
For a first estimate of the effect we
can make use of the effective radii listed in Table~\ref{table parameters}:
these give an indication of the extent of the radio photospheres.
We see that for the shorter wavelengths, even at periastron 
(where the separation between the stars is $\sim$ 1300~R$_{\sun}$)
the CWR is 
outside the radio photospheres of each of the stars.
Therefore no significant absorption
of the synchrotron photons occurs.

Away from periastron, the spectral index is nearly flat, or slightly
negative. The fluxes at all wavelengths are of order 2~mJy, slightly
lower at 2~cm and somewhat higher at 20~cm. 
This indicates that we are
seeing the non-thermally emitting CWR during a large part of the orbit.
It is therefore not completely hidden part of the time, as was the
case during the periastron passage of 
Cyg~OB2\,\#~9 \citep{2013A&A...550A..90B}. As mentioned above,
this can be attributed to the longer
period of 9~Sgr, resulting in a CWR region that stays out of the free-free
region of each star.

The contribution of the thermal free-free emission
of the stellar winds of both
stars is negligible. Using the equations of
\citet{1975MNRAS.170...41W} we can calculate the expected 
thermal radio flux,
using the parameters listed in Table~\ref{table parameters}.
The combined flux of both binary components is 0.02~mJy at 2~cm and
0.005~mJy at 20~cm, which is much smaller than the observed fluxes.

\section{Modelling}
\label{section modelling}

\begin{table}
\caption{Parameters of 9~Sgr used in this study.}
\label{table parameters}
\centering
\begin{tabular}{llcc}
\hline
\hline
Parameter                  & symbol & primary & secondary \\
\hline
orbital period             & $P$ (d)                   & \multicolumn{2}{c}{$3327$} \\
time periastron passage & $T_0$ (JD)                & \multicolumn{2}{c}{$2\,446\,564$} \\
eccentricity               & $e$                       & \multicolumn{2}{c}{$0.702 \pm 0.010$} \\
longitude of periastron    & $\omega$\,(\degr)          & \multicolumn{2}{c}{$27.4 \pm 1.2$} \\
major axis                 & $a \sin i$\,($R_{\sun}$)     & \multicolumn{2}{c}{$3011$} \\
mass                       & $m \sin^3 i$\,($M_{\sun}$) & $20.0$ & $13.3$ \\
mass-loss rate             & $\dot{M}\,(M_{\sun}\,{\rm yr}^{-1}$)  & $9.0\times 10^{-7}$ & $5.0\times 10^{-7}$ \\
terminal velocity          & $v_{\infty}\,({\rm km\,s}^{-1}$)   & 3500 & 3100 \\
distance                   & $D$ (kpc)                   & \multicolumn{2}{c}{1.79} \\
estimated inclination      & $i$ (\degr)                 & \multicolumn{2}{c}{45} \\
thermal flux at 2 cm       & $F_{\nu,2}$ (mJy)           & 0.014 & 0.007 \\
thermal flux at 6 cm       & $F_{\nu,6}$ (mJy)           & 0.007 & 0.004 \\
thermal flux at 20 cm       & $F_{\nu,20}$ (mJy)         & 0.003 & 0.002 \\
effective radius, 2 cm    & $R_{\rm eff,2}$ ($R_{\sun}$) &  260 & 190 \\
effective radius, 6 cm    & $R_{\rm eff,6}$ ($R_{\sun}$) &  560 & 410 \\
effective radius, 20 cm   & $R_{\rm eff,20}$ ($R_{\sun}$) & 1300 & 950 \\
\hline
\end{tabular}
\tablefoot{Orbital parameters from Rauw (2013, pers. comm.), which is an updated version of the
results published by \citet{2012A&A...542A..95R}. Radio data are determined
from the \citet{1975MNRAS.170...41W} equations.}
\end{table}

To model the radio-light curve, we basically use the same simple model
as presented by \citet{2013A&A...550A..90B}. 
The use of such a simple model avoids the introduction of less well-known
quantities, such as the shock strength, the local magnetic field and the
efficiency of Fermi mechanism.

One of the simplifications in the model is that it assumes a simple 
shape for the CWR. This is more consistent with a CWR in the adiabatic
regime than a radiative one where the instabilities create a lot
of structure \citep{1992ApJ...386..265S, 2009MNRAS.396.1743P}.
That the CWR of 9~Sgr stays in the adiabatic regime 
can be seen by comparing the
cooling time ($t_{\rm cool}$) to the escape time ($t_{\rm esc}$),
using the equation from \citet{1992ApJ...386..265S}:
\begin{equation}
\chi = \frac{t_{\rm cool}}{t_{\rm esc}} \approx \frac{v_8^4 d_{12}}{\dot{M}_{-7}},
\end{equation}
with $v_8$ the wind velocity in units of 1000 km\,s$^{\rm -1}$, $d_{12}$ the 
distance to the
contact discontinuity in units of $10^7$ km and $\dot{M}_{-7}$ the mass-loss 
rate in units of $10^{-7}$ $M_{\sun}\,{\rm yr}^{-1}$. 
Using the values from Table~\ref{table parameters}
and assuming the contact discontinuity to be half-way between the two stars,
we find $\chi \approx 740 - 4\,700$, where the range covers the difference between
periastron and apastron values. The high $\chi$ values indicate that
the collision is indeed adiabatic \citep[see also][]{2012A&A...542A..95R}.

In the model, we solve the radiative transfer
in a three-dimensional grid. For each phase in the orbit, we position the
stars in this grid, taking into account the orbital parameters and
the estimated inclination 
(Table~\ref{table parameters}). For a large part of the grid,
the volume is filled with wind material coming from the closer-by star.
This wind material will contribute to the radio flux through the
free-free emission (but we expect the contribution to be small --
see Sect.~\ref{section radio light curve}).
The density at any point in the model
is derived from the mass-loss rate and terminal
velocity (Table~\ref{table parameters}). 
We caution that these two quantities are not well known. The spectral
classification of both components is hampered by the fact that the
He {\sc ii} lines cannot be well disentangled \citep{2012A&A...542A..95R}.
This classification is then used to determine the 
mass-loss rate and terminal
velocity from the \citet{2005A&A...436.1049M} calibration. Furthermore,
any given spectral type/luminosity class corresponds to a range
of effective temperature and luminosity 
rather than the unique values given by the
\citeauthor{2005A&A...436.1049M} calibration \citep{2010A&A...524A..98W}.
This turned out to be an important effect in our analysis
of Cyg~OB2\,\#9 \citep{2013A&A...550A..90B}.

For material inside the CWR, we multiply the
density by a factor of 4 to take into account the compression in the shocks.
The wind material is assumed to be at T=20,000 K, i.e.
just below half of the effective temperature of the stars.
A higher temperature is assigned to the material inside the
CWR (see below).
The grid is $100,000~{\rm R_{\sun}}$ on each side and is centred on the
centre of gravity of the binary system. 

\begin{figure}
\centering
\resizebox{\hsize}{!}{\includegraphics[bb=24 32 275 322]{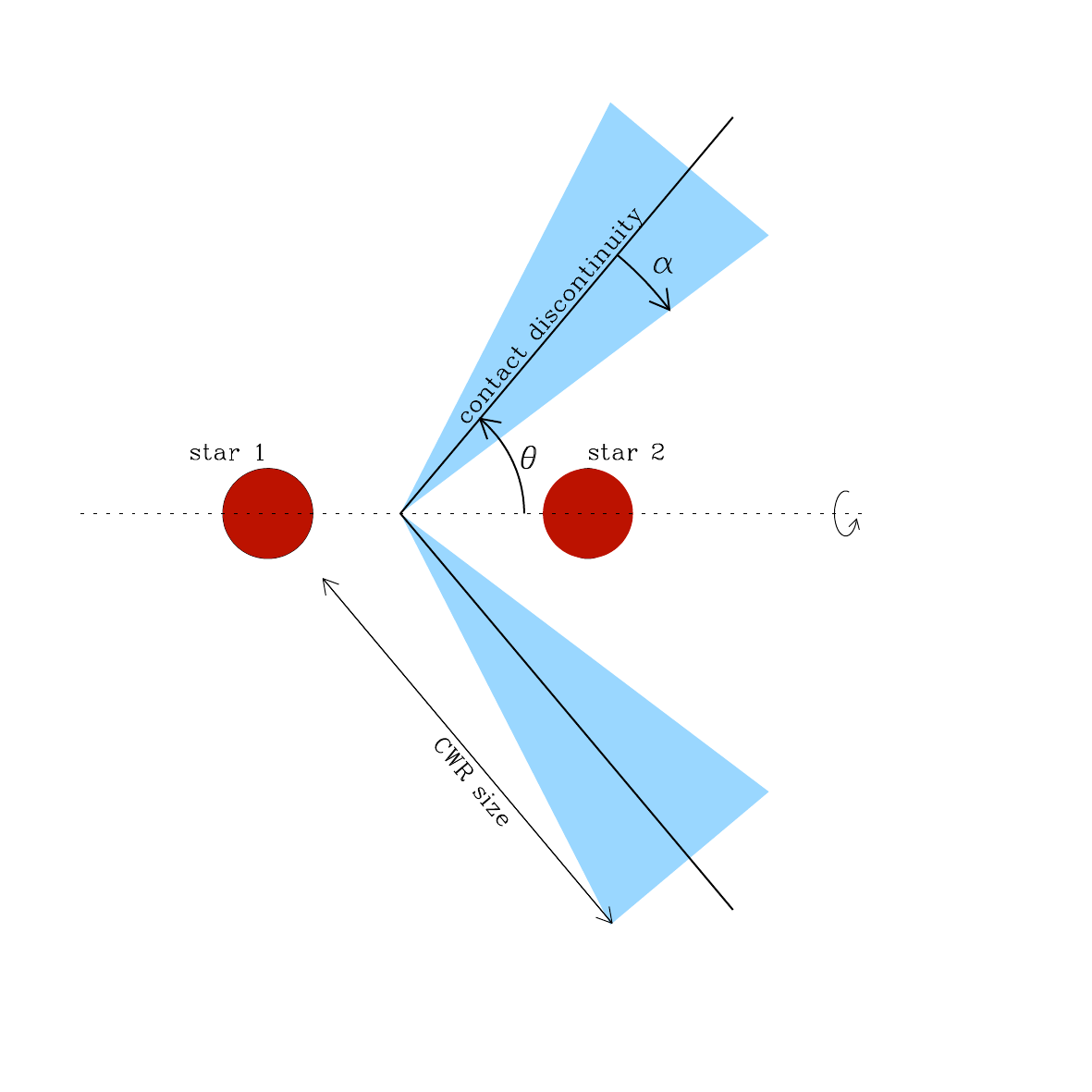}}
\caption{Schematic view of the second option for our model for the CWR. 
At any given phase, the shape of the CWR (shaded in light-blue)
is a cone that is rotationally symmetric around the axis connecting the
two stars. It has a (half) opening angle $\theta$, a
(half) flaring angle $\alpha$ and a size ($R_{\rm CWR}$) which scales
with the separation ($D$) between the two components.}
\label{fig CWR}
\end{figure}

For the shape of the
CWR, we considered two options. The first one is as
presented in \citet{2013A&A...550A..90B}: the region has the shape of
a cone, rotationally symmetric around the line connecting the two stars.
The position of this cone and its opening angle are derived from
\citet[][their Eqs. (1) and (3)]{1993ApJ...402..271E}. The cone is 
also given a finite thickness which remains constant in the whole
simulation volume
(\citeauthor[][their Fig.~3]{2013A&A...550A..90B}).
Hydrodynamical simulations of adiabatic wind
collisions
\citep[e.g.][their Fig.~4]{2011MNRAS.418.2618L}
show a somewhat different shape. In these calculations,
the CWR is thin near the apex (on the line connecting the 
two stars) and flares out as we move away from that line.
We therefore also considered a second option, where the cone has a 
``flaring" angle ($\alpha$) instead of a finite thickness.
(see Fig.~\ref{fig CWR}). For both options, we limit the size of the
CWR. \cite{1993ApJ...402..271E} estimate the size to be $\pi$ times the
distance from the apex to the weaker-wind star. In our model,
we assume that the size (denoted $R_{\rm CWR}$)
scales proportionally to the distance between 
the two stars ($D$), with the scaling factor a free parameter.

Because the material inside the CWR is hot and compressed it will
also contribute to the radio flux through free-free emission
\citep{2010MNRAS.403.1633P}.
We therefore need to assign a temperature to this hot material. 
\citet{2002A&A...394..993R} fit the X-ray
emission of this system with a multi-temperature thermal model
with components at $kT$=0.25--0.26~keV, 0.67--0.73~keV and 
possibly $>1,46$~keV.
This corresponds to temperatures of $3 \times 10^{6}$, $8 \times 10^{6}$, 
and $2 \times 10^{7}$~K respectively.
In the models we typically run, we calculate the emission measure for the
CWR material as:
\begin{equation}
{\rm EM} = \int_{\rm CWR} n_{\rm e} n_{\rm H} {\rm d}V,
\end{equation}
where $n_{\rm e}$ is the electron number density and $n_{\rm H}$ is the proton
number density. The integration is over the whole volume of the CWR,
and gives a ranges of values 
of $\log {\rm EM} \approx 55.2-57.2$ (EM in ${\rm cm}^{-3}$).
The observed emission measures of the X-ray temperatures are respectively
$\log {\rm EM}= 56.2$, $55.5$ and $54.7$. This leads to an average temperature
of the model CWR of $2 \times 10^5 - 5 \times 10^6$~K. To simplify the model,
we assign a single temperature of $1.0 \times 10^6$~K to the hot CWR
material. Because of the high ratio of
the cooling time to the escape time, these temperatures will not 
significantly decrease inside the CWR
as we move away from the apex.

We further assume that all the material in the CWR also emits
synchrotron radiation. The amount emitted is described by its brightness 
temperature ($T_{\rm sync}$), 
which we consider as a free parameter in our model.

The radiative transfer equation
is then solved in this 3D grid, following the procedure
outlined in \cite{1975MNRAS.170...41W}.
We take into account
the free-free emission and absorption processes in the stellar winds and CWR,
as well as the synchrotron emission in the CWR. 
Note that the synchrotron brightness
temperature does not play a role in the absorption
because synchrotron self-absorption has little effect
at GHz frequencies \citep{2006A&A...446.1001P}.
We use an adaptive grid
scheme, which refines the grid only in those places where needed 
to get the required
precision in specific intensity and flux.

\section{Results and discussion}
\label{section discussion}

\begin{figure}
\resizebox{\hsize}{!}{\includegraphics{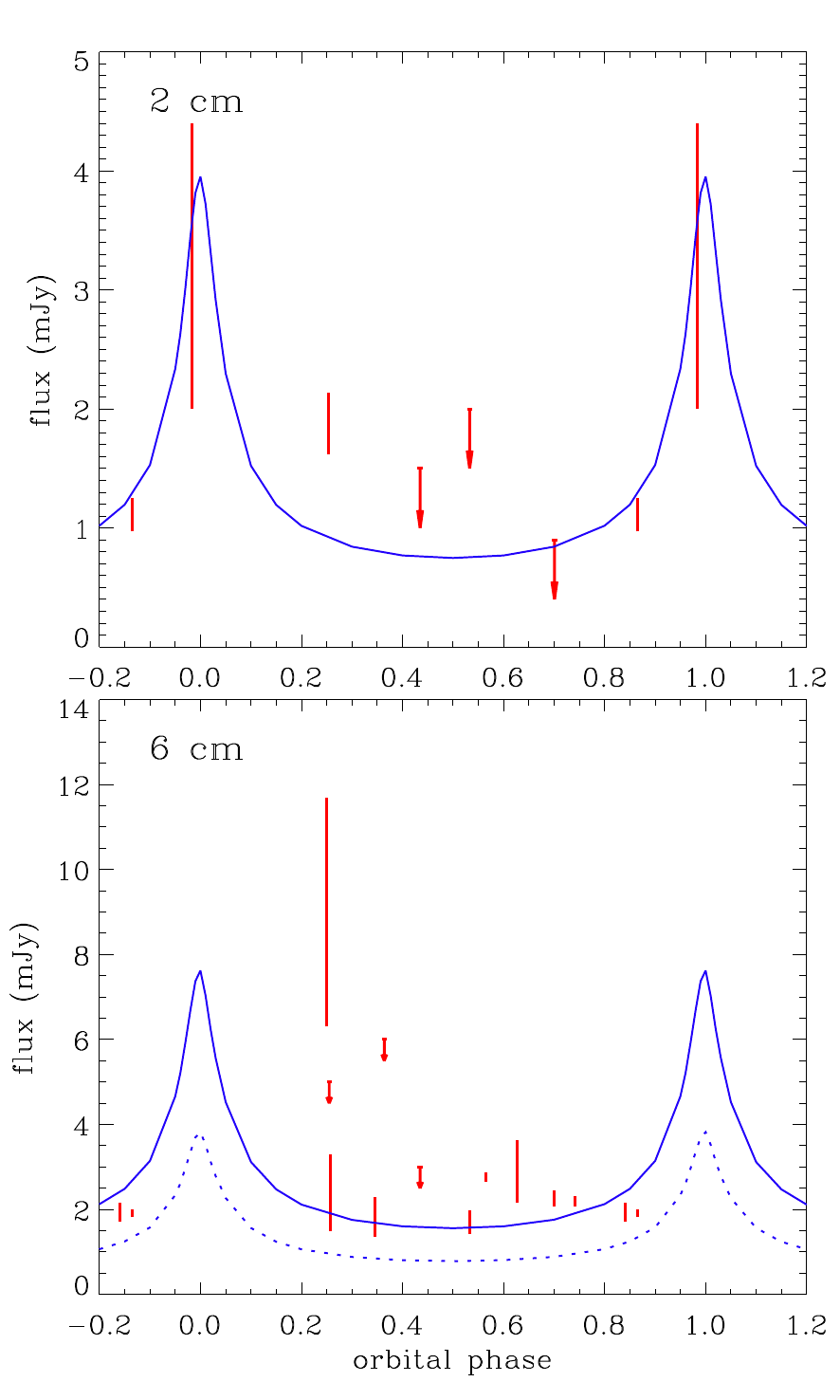}}
\caption{Comparison between 9~Sgr observed and model radio fluxes, 
as a function of orbital phase in the 9.1-year period. Top: 2 cm
fluxes, bottom: 6 cm fluxes. The observed data are plotted in red, the
theoretical fluxes of the best-fit model
in blue. 
In the bottom panel, the dotted blue line
show the best-fit 2-cm model applied to the 6-cm data.
}
\label{fig model fluxes}
\end{figure}

We started by fitting the 2-cm radio light curve, because at this
wavelength we see the clearest evidence of phase-locked variability
(see Fig.~\ref{fig fluxes}). For the observed light curve we used
the fluxes from the ``combined" data (see Sect.~\ref{section data reduction})
instead of the separate ones, because the combined data
have a smaller error bar.

In the model calculations we tried both options for the shape
of the CWR, but the more realistic one -- where the thickness of the CWR
increases as we move away from the apex -- gives the better results.
We explored the parameter space of the synchrotron brightness 
temperature ($T_{\rm sync}$), flaring angle ($\alpha$)
and size of the CWR ($R_{\rm CWR}$). 
Formally, the best-fit
parameters are: $T_{\rm sync} = 4.0 \times 10^8$~K, $\alpha = 40$\degr,
$R_{\rm CWR} = 2.0~D$; this solution
is shown on Fig.~\ref{fig model fluxes} (top panel, blue solid line).
The result is not very unique however. It is quite insensitive to the
size of the CWR, and other combinations of $T_{\rm sync}$
and $\alpha$ can give an almost equally good fit. Our calculations
show that this is the case when the
combination $T_{\rm sync} \alpha^{0.8}$ has the same value as
our formal best fit. Values for
the half-flaring angle larger than 40\degr~were not explored, as this
would make the size of the CWR hydrodynamically unlikely.

The agreement of the theoretical curve with the observed fluxes is good.
We correctly predict an increase of the 2-cm flux around periastron
(up to $\sim$~4 mJy)
and a roughly constant flux away from periastron (at the $\sim$~1 mJy level).
The only discrepant point is at phase 0.25. This indicates that the 
light curve around periastron is not as symmetric as suggested by the
theoretical calculations: the higher flux level extends for a longer
time after periastron passage than before it.

We explored the contribution of the various emission and absorption
processes to this theoretical light curve by manipulating the temperature
and density enhancement in the CWR. If we switch off the synchrotron emission
and the density enhancement in the CWR, and set the CWR temperature equal
to the wind temperature, we find the free-free contribution of the stellar
winds. This turns out to be 0.020 mJy, in good agreement with the 
values listed in Table~\ref{table parameters}. By switching off only the
synchrotron emission, we find the combined free-free contribution of
the stellar winds and the CWR: this is 0.021--0.027 mJy (depending
on orbital phase). The free-free emission of the CWR is therefore
very small.
Finally, switching off the free-free absorption
in the stellar wind changes the fluxes by only a few percent.
The 2-cm radio light curve
in Fig.~\ref{fig model fluxes} is therefore almost exclusively due to
synchrotron emission. 

Free-free absorption in the CWR does play an important 
role, as can be seen from the following argument. An object with a
brightness temperature $T_{\rm sync}=4.0 \times 10^8$~K at a 
distance $d=1.79$~kpc
with a size equal to the separation between the two components 
($D=3011/\sin(45\degr)$~${\rm R}_{\sun}$) has a flux of
\begin{equation}
S_\nu = \pi \left(\frac{D}{d}\right)^2 \frac{2 k T_{\rm sync}}{\lambda^2} \approx 25\,000~{\rm mJy}.
\end{equation}
This is clearly more than the 1--4 mJy from the model calculation, hence a
large amount of the synchrotron flux must be absorbed in the CWR.

The model used here has a number of simplifying assumptions. One is that
the synchrotron brightness temperature is the same for all phases in the
orbit. The synchrotron emission is however proportional to the
number of relativistic electrons \citep[][Chap.~6]{1986rpa..book.....R}.
We therefore also tried models with a synchrotron brightness temperature
that is proportional to the local electron density (this assumes that 
the fraction of electrons that becomes relativistic is the same at 
all phases of the orbit). Such models result in a much higher contrast between
the maximum and minimum flux, and are therefore not a good fit to the
2-cm observations.

We next applied the same model to the 6~cm fluxes. This is shown by the
dotted blue line in the bottom panel of Fig.~\ref{fig model fluxes}. The
model clearly underestimates the observed fluxes. This is not surprising
as the synchrotron brightness temperature need not be the same for 6 cm
as for 2 cm. In a more detailed synchrotron emission model 
\citep[such as used by][]{2010A&A...519A.111B}
we would, in principle, be able to calculate the synchrotron brightness 
temperature for 
all wavelengths. We note however that the model of 
\citeauthor{2010A&A...519A.111B} was not able to explain the observed
spectral index of Cyg~OB2\,\#8A, a quantity which is directly related to
the relative synchrotron brightness temperatures. 

We therefore again considered the synchrotron brightness temperature as
a free parameter and tried to fit the 6-cm observations. We kept the
flaring angle and CWR size at their 2-cm best-fit values. The best-fit
result is shown by the full blue line in Fig.~\ref{fig model fluxes} (bottom
panel) and has a
brightness temperature of $T_{\rm sync}=8.0 \times 10^8$~K.
Away from periastron we obtain the $\sim$~2 mJy level, as observed.
We have no 6-cm observations around periastron itself, so we cannot judge
if the flux maximum has the correct value. The most discrepant point in the
fit is what happens around phase 0.25: there we have two observations which
are seemingly in contradiction: one at $9.0\pm 2.7$~mJy and one at 
$2.4 \pm 0.9$~mJy. 
These flux determinations do not suffer from the measurement
problems due to the nearby Hourglass Nebula
(Sect.~\ref{section data reduction}). They were taken in VLA
configurations that have medium to high spatial resolution,
where the effect of the Hourglass Nebula is small.
We note however that both have large error bars,
so these detection are only at the $\sim$~$3\sigma$ level.
As for the 2-cm light curve, the parameters are not well determined
and other combinations give fits of a similar quality.

9~Sgr is exceptional among the O-type non-thermal radio emitters in that
it is the one with the longest period among those that have their
spectroscopic orbit determined. Cyg~OB2\,\#8A has a substantially shorter
period (21.9~d). While one might expect that in such a short-period
binary all synchrotron emission would be absorbed by the stellar winds,
it turns out that this is still a non-thermal radio emitter
\citep{2004A&A...424L..39D, 2010A&A...519A.111B}. 
HD~168112 \citep[$P=1.4?$ yr,][]{2005A&A...436.1033B} and Cyg~OB2\,\#9 
\citep[$P=2.35$ yr,][]{2008A&A...483..543N,2008A&A...483..585V,2012A&A...546A..37N,2013A&A...550A..90B} 
are longer-period binaries with clear non-thermal emission. For Cyg~OB2\,\#9
we know that the CWR is radiative at least during some part of the orbit.
This complicates the modelling of the emission. Furthermore, with
these periods, the effect of orbital motion that turns the CWR into
a spiral shape, will be important \citep[e.g.][]{2009MNRAS.396.1743P}.

There are other O-type non-thermal radio emitters that have longer periods.
\object{Cyg~OB2\,\#5} 
is a 6.6-d spectroscopic binary, but the 6.7-yr period in the
radio fluxes indicates the presence of a third companion. Furthermore,
a CWR is visible at radio wavelengths between Cyg~OB2\,\#5 and a nearby B-type
star, implying that Cyg~OB2\,\#5 could be a quadruple system
\citep{2010ApJ...709..632K, 2013ApJ...763..139D}. From spectroscopy we
know that HD~167971 is a 3.3-d binary, with a third component which may
or may not be gravitationally bound \citep{1987A&A...185..121L}.
Radio data do not show the 3.3-d period, but reveal a period of $\sim$~20~yr,
or longer, suggesting that this is due to the CWR between the binary and
the third component \citep{2007A&A...464..701B}.

What makes 9~Sgr an excellent candidate for modelling studies is that it
has an adiabatic wind collision and that it is much less influenced by
orbital motion. Longer-period systems (such as Cyg~OB2\,\#5 and HD~167971)
might be even better, but they lack the spectroscopic orbital information.
A disadvantage of 9~Sgr is that the radio coverage of the orbit, especially
at periastron, is somewhat lacking. It should however be easy to remedy that 
situation thanks to the recent upgrade of radio telescopes such as the
Karl G. Jansky Very Large Array (JVLA).

\section{Conclusions}
\label{section conclusions}

Using archive data from the VLA that cover a time range of 24 years, 
we determined the radio light curve
of the massive O-type binary 9~Sgr. The presence of the nearby Hourglass
Nebula, a strong radio source, seriously hampers the detection of
9~Sgr when the VLA is in one of its configurations with lower spatial
resolution. The quality of the radio light curve is therefore less than
that of other systems studied in this series of papers.

Notwithstanding these problems, the 2-cm light curve shows clear
phase-locked variability with the 9.1-yr orbit of this system. 
Fluxes are higher around periastron, as expected, because in this
highly eccentric system the wind-wind collision is much stronger (i.e.
has a higher ram pressure) when the stars are closer to each other.
The 6~cm light curve seems to follow a similar trend, although observations
at the periastron passage are missing. The few data we have at 20~cm
are more puzzling, as they seem to show large variations at other phases
than periastron. The spectral index is approximately zero, and even outside
the periastron phases the fluxes are so high that almost all of the emission
can be ascribed to synchrotron radiation in the colliding-wind region (CWR).

A simple model provides a good fit to the 2-cm observations, and allows
us to estimate that the synchrotron brightness
temperature of the CWR is at least $4 \times 10^8$~K. 
Higher values of the brightness temperature are also possible, 
provided the flaring angle is smaller. The geometric
extent of the CWR is not well constrained in this fitting procedure.
We caution that this
simple model lacks many important ingredients, such as the hydrodynamics
of the CWR, the efficiency with which the electrons get accelerated up
to relativistic speeds, the local magnetic field and the quenching due
to the Razin effect.

We propose that 9~Sgr is an ideal candidate for more detailed modelling
of its radio emission. Its orbital parameters are sufficiently well known.
It has a long period and because of the larger distance between the two
components the collision region remains adiabatic during the whole orbit.
Also, the effect of orbital motion on the shape of the CWR is much smaller
than in other binary systems. All
this simplifies the hydrodynamical part of the modelling. Observationally,
the coverage of the orbit should be improved, especially around periastron.
This should be quite feasible using recently upgraded radio telescopes
such as the JVLA.

\begin{acknowledgements}
We thank Gregor Rauw for providing updated values for the orbital
parameters and Joan Vandekerckhove for his help with the reduction of the
VLA data. We are grateful to
the original observers of the VLA archive data used in this paper.
D. Volpi acknowledges funding by the Belgian Federal Science Policy Office 
(Belspo), under contract MO/33/024.
This research has made use of the SIMBAD database, operated at CDS,
Strasbourg, France and NASA's Astrophysics Data System Abstract Service.
\end{acknowledgements}

\bibliographystyle{aa}
\bibliography{9Sgr}

\Online
\appendix

\section{Data table}

\begin{longtable}{llrrcrrrlcl}
\caption{\label{table radio data}9~Sgr VLA data. 
}\\
\hline\hline
\multicolumn{1}{c}{(1)} & \multicolumn{1}{c}{(2)} & \multicolumn{1}{c}{(3)} &
\multicolumn{1}{c}{(4)} & \multicolumn{1}{c}{(5)} & \multicolumn{1}{c}{(6)} &
\multicolumn{1}{c}{(7)} & \multicolumn{1}{c}{(8)} & \multicolumn{1}{c}{(9)} &
\multicolumn{1}{c}{(10)} \\
\multicolumn{1}{c}{Progr.} & \multicolumn{1}{c}{Date} &
\multicolumn{1}{c}{Ctr.} &
\multicolumn{3}{c}{Phase Calibrator} &
\multicolumn{1}{c}{Int.} &
\multicolumn{1}{c}{No.} &
\multicolumn{1}{c}{Config.} &
\multicolumn{1}{c}{Flux} \\
\cline{4-6}
&  & &
\multicolumn{1}{c}{Name} &
\multicolumn{1}{c}{Flux} &
\multicolumn{1}{c}{Dist.} &
\multicolumn{1}{c}{Time} &
\multicolumn{1}{c}{Ants.} & &
\multicolumn{1}{c}{} & \\
&  & &
\multicolumn{1}{c}{} &
\multicolumn{1}{c}{(Jy)} &
\multicolumn{1}{c}{(degr.)} &
\multicolumn{1}{c}{(min)} &
\multicolumn{1}{c}{} & &
\multicolumn{1}{c}{(mJy)} & \\
\hline
\endfirsthead
\caption{continued.}\\
\hline\hline
\multicolumn{1}{c}{(1)} & \multicolumn{1}{c}{(2)} & \multicolumn{1}{c}{(3)} &
\multicolumn{1}{c}{(4)} & \multicolumn{1}{c}{(5)} & \multicolumn{1}{c}{(6)} &
\multicolumn{1}{c}{(7)} & \multicolumn{1}{c}{(8)} & \multicolumn{1}{c}{(9)} &
\multicolumn{1}{c}{(10)} \\
\multicolumn{1}{c}{Progr.} & \multicolumn{1}{c}{Date} &
\multicolumn{1}{c}{Ctr.} &
\multicolumn{3}{c}{Phase Calibrator} &
\multicolumn{1}{c}{Int.} &
\multicolumn{1}{c}{No.} &
\multicolumn{1}{c}{Config.} &
\multicolumn{1}{c}{Flux} \\
\cline{4-6}
&  & &
\multicolumn{1}{c}{Name} &
\multicolumn{1}{c}{Flux} &
\multicolumn{1}{c}{Dist.} &
\multicolumn{1}{c}{Time} &
\multicolumn{1}{c}{Ants.} & &
\multicolumn{1}{c}{} & \\
&  & &
\multicolumn{1}{c}{} &
\multicolumn{1}{c}{(Jy)} &
\multicolumn{1}{c}{(degr.)} &
\multicolumn{1}{c}{(min)} &
\multicolumn{1}{c}{} & &
\multicolumn{1}{c}{(mJy)} & \\
\hline
\endhead
\hline
\endfoot
\endlastfoot
\\\multicolumn{11}{c}{{\bf 0.7 cm}} \\
\hline
  AR328 & 1995-04-27 &       9 Sgr &   1733-130 & $   10.67 \pm     0.08$ & $ 13.4$ & $   17$ &    9 &    D & $<        25.$  \\
\hline
\\\multicolumn{11}{c}{{\bf 2 cm}} \\
\hline
   BIEG & 1982-02-09 &       9 Sgr &   1733-130 & $    4.17 \pm     0.05$ & $ 13.4$ & $  103$ &   23 &   AD & $<         2.$  \\
   AC42 & 1983-08-22 &       9 Sgr &   1733-130 & $    5.91 \pm     0.10$ & $ 13.4$ & $   47$ &   23 &    A & $<        0.9$  \\
  AB327 & 1985-01-28 &       9 Sgr &   1733-130 & $    6.99 \pm     0.14$ & $ 13.4$ & $   30$ &   26 &    A & $   1.46 \pm    0.30$  \\
  AC116 & 1985-02-16 &       9 Sgr &   1733-130 & $    6.46 \pm     0.04$ & $ 13.4$ & $   37$ &   24 &    A & $   1.01 \pm    0.28$  \\
\multicolumn{4}{l}{{\em combined     AB327+AC116 }}            &                         &         &         &      &      & $   1.11 \pm    0.14$  \\
   AT89 & 1988-01-22 &          M8 &   1733-130 & $    5.16 \pm     0.03$ & $ 13.4$ & $  105$ &   25 &    B & $<        15.$  \\
  AH265 & 1988-05-05 &       M8UU1 &   1733-130 & $    4.92 \pm     0.08$ & $ 13.4$ & $   40$ &   22 &   CD & $<        50.$  \\
  AR328 & 1995-04-27 &       9 Sgr &   1733-130 & $    9.74 \pm     0.03$ & $ 13.4$ & $   22$ &   15 &    D & $(    3.2 \pm     1.2)$  \\
  AH557 & 1996-01-23 &          M8 &   1733-130 & $   11.10 \pm     0.05$ & $ 13.4$ & $   33$ &   26 &  CnB & $<        50.$  \\
  AW478 & 1997-10-03 &       9 Sgr &   1744-312 & $   0.751 \pm    0.007$ & $  8.1$ & $   13$ &   22 &  DnC & $<        16.$  \\
  AW478 & 1997-10-07 &       9 Sgr &   1744-312 & $   0.788 \pm    0.002$ & $  8.1$ & $   13$ &   25 &  DnC & $    2.6 \pm     0.9$  \\
  AW478 & 1997-10-11 &       9 Sgr &   1744-312 & $   0.790 \pm    0.007$ & $  8.1$ & $   13$ &   25 &  DnC & $<         4.$  \\
  AW478 & 1997-10-15 &       9 Sgr &   1744-312 & $   0.844 \pm    0.003$ & $  8.1$ & $   13$ &   26 &  DnC & $<         2.$  \\
\multicolumn{4}{l}{{\em combined           AW478 }}            &                         &         &         &      &      & $   1.88 \pm    0.26$  \\
  AW515 & 1999-06-08 &       9 Sgr &   1820-254 & $   0.896 \pm    0.003$ & $  4.0$ & $   30$ &   18 &   AD & $<        1.5$  \\
\hline
\\\multicolumn{11}{c}{{\bf 3.6 cm}} \\
\hline
  AW304 & 1992-01-24 &       9 Sgr &   1924-292 & $   17.13 \pm     0.78$ & $ 18.7$ & $   11$ &   24 &  CnB & $<         9.$  \\
  AB671 & 1993-01-21 &       9 Sgr &   1811-209 & $  0.1684 \pm   0.0007$ & $  3.8$ & $   19$ &   22 &    A & $   1.59 \pm    0.09$  \\
  AB671 & 1993-01-24 &       9 Sgr &   1811-209 & $  0.1709 \pm   0.0005$ & $  3.8$ & $   19$ &   26 &    A & $   1.41 \pm    0.13$  \\
  AB671 & 1993-01-29 &       9 Sgr &   1811-209 & $  0.1713 \pm   0.0004$ & $  3.8$ & $   19$ &   25 &  BnA & $   1.71 \pm    0.26$  \\
  AB671 & 1993-02-01 &       9 Sgr &   1811-209 & $  0.1699 \pm   0.0006$ & $  3.8$ & $   13$ &   26 &  BnA & $   1.33 \pm    0.18$  \\
  AB671 & 1993-02-14 &       9 Sgr &   1811-209 & $  0.1745 \pm   0.0004$ & $  3.8$ & $    9$ &   26 &  BnA & $   1.80 \pm    0.30$  \\
\multicolumn{4}{l}{{\em combined           AB671 }}            &                         &         &         &      &      & $  1.36 \pm   0.06$  \\
  AR328 & 1995-04-27 &       9 Sgr &   1733-130 & $    6.74 \pm     0.02$ & $ 13.4$ & $   11$ &   15 &    D & $<         6.$  \\
  AH557 & 1995-09-25 &          M8 &   1733-130 & $   7.879 \pm    0.015$ & $ 13.4$ & $   28$ &   26 &  BnA & $   3.72 \pm    0.44$  \\
  AW478 & 1997-10-03 &       9 Sgr &   1751-253 & $  0.2579 \pm   0.0008$ & $  2.9$ & $   13$ &   26 &  DnC & $<         3.$  \\
  AW478 & 1997-10-07 &       9 Sgr &   1751-253 & $  0.2724 \pm   0.0003$ & $  2.9$ & $   13$ &   26 &  DnC & $<         2.$  \\
  AW478 & 1997-10-11 &       9 Sgr &   1751-253 & $  0.2665 \pm   0.0006$ & $  2.9$ & $   13$ &   26 &  DnC & $<         4.$  \\
  AW478 & 1997-10-15 &       9 Sgr &   1751-253 & $  0.2698 \pm   0.0004$ & $  2.9$ & $   13$ &   26 &  DnC & $<        90.$  \\
  AW478 & 1997-10-18 &       9 Sgr &   1751-253 & $  0.2710 \pm   0.0005$ & $  2.9$ & $    9$ &   25 &  DnC & $    3.8 \pm     1.3$  \\
\multicolumn{4}{l}{{\em combined           AW478 }}            &                         &         &         &      &      & $   1.78 \pm    0.36$  \\
 AB1005 & 2001-03-08 &       9 Sgr &   1751-253 & $  0.2658 \pm   0.0015$ & $  2.9$ & $    8$ &   25 &    B & $    3.6 \pm     0.9$  \\
  AW574 & 2002-03-29 &         [1] &   1751-253 & $  0.2705 \pm   0.0005$ & $  2.9$ & $    2$ &   25 &    A & $<         5.$  \\
  AK559 & 2003-04-05 &       G9.97 &   1820-254 & $  0.6580 \pm   0.0008$ & $  4.1$ & $   15$ &   25 &    D & $<        20.$  \\
  AF399 & 2003-06-07 &         [2] &   1820-254 & $   0.636 \pm    0.002$ & $  4.0$ & $   21$ &   27 &    A & $  1.81 \pm   0.08$  \\
 AB1094 & 2003-07-05 &       18006 &   1820-254 & $  0.7611 \pm   0.0012$ & $  4.1$ & $   20$ &   26 &    A & $    2.3 \pm    0.55$  \\
  AF399 & 2003-09-09 &         [2] &   1820-254 & $  0.6825 \pm   0.0019$ & $  4.0$ & $   21$ &   27 &    A & $   1.85 \pm    0.17$  \\
\hline
\\\multicolumn{11}{c}{{\bf 6 cm}} \\
\hline
   CHUR & 1979-07-13 &       9 Sgr &   1733-130 & $    5.38 \pm     0.10$ & $ 13.4$ & $  186$ &   13 &   AC & $    9.0 \pm     2.7$  \\
   NEWE & 1979-08-07 &       9 Sgr &   1733-130 & $    5.64 \pm     0.13$ & $ 13.4$ & $  112$ &   14 &   AC & $(    5.2 \pm     2.4)$  \\
   NEWE & 1979-08-09 &       9 Sgr &   1733-130 & $    5.38 \pm     0.10$ & $ 13.4$ & $   30$ &   14 &   AC & $<        20.$  \\
\multicolumn{4}{l}{{\em combined            NEWE }}            &                         &         &         &      &      & $(    2.4 \pm     0.9)$  \\
   CHUR & 1980-05-23 &       9 Sgr &   1733-130 & $    4.60 \pm     0.10$ & $ 13.4$ & $   30$ &   21 &   AC & $<         4.$  \\
   CHUR & 1980-05-24 &       9 Sgr &   1733-130 & $    5.40 \pm     0.02$ & $ 13.4$ & $   49$ &   22 &   AC & $   1.88 \pm    0.40$  \\
\multicolumn{4}{l}{{\em combined            CHUR }}            &                         &         &         &      &      & $   1.82 \pm    0.47$  \\
   BIEG & 1980-07-26 &       9 Sgr &   1733-130 & $   5.242 \pm    0.015$ & $ 13.4$ & $   10$ &   20 &    C & $<         8.$  \\
   BIEG & 1980-07-27 &       9 Sgr &   1733-130 & $   5.187 \pm    0.017$ & $ 13.4$ & $   19$ &   20 &    C & $<        11.$  \\
\multicolumn{4}{l}{{\em combined            BIEG }}            &                         &         &         &      &      & $<         6.$  \\
   SIMO & 1980-09-08 &      Her 36 &   1911-201 & $    2.51 \pm     0.03$ & $ 16.2$ & $    0$ &   23 &    D & $<         9.$  \\
   BROW & 1981-12-28 &      G5.973 &   1751-253 & $   0.532 \pm    0.002$ & $  2.9$ & $    6$ &   26 &    C & $<         9.$  \\
   BIEG & 1982-02-09 &       9 Sgr &   1733-130 & $    4.82 \pm     0.04$ & $ 13.4$ & $   19$ &   23 &   AC & $   1.70 \pm    0.28$  \\
   BIGN & 1982-05-26 &       9 Sgr &   1733-130 & $   5.090 \pm    0.013$ & $ 13.4$ & $   34$ &   25 &    A & $   2.76 \pm    0.11$  \\
   BROW & 1982-11-12 &      G5.973 &   1751-253 & $   0.534 \pm    0.004$ & $  2.9$ & $    6$ &   26 &    D & $<        25.$  \\
   AH11 & 1983-02-06 &          M8 &   1911-201 & $   2.668 \pm    0.015$ & $ 16.2$ & $   50$ &   25 &    C & $<         9.$  \\
   AJ97 & 1983-08-19 &       9 Sgr &   1733-130 & $    5.47 \pm     0.18$ & $ 13.4$ & $   55$ &   25 &    A & $   2.35 \pm    0.32$  \\
   AC42 & 1983-08-22 &       9 Sgr &   1733-130 & $   5.090 \pm    0.015$ & $ 13.4$ & $   32$ &   24 &    A & $   2.20 \pm    0.14$  \\
\multicolumn{4}{l}{{\em combined       AJ97+AC42 }}            &                         &         &         &      &      & $   2.27 \pm    0.19$  \\
  AC116 & 1984-11-27 &       9 Sgr &   1733-130 & $   4.978 \pm    0.006$ & $ 13.4$ & $    9$ &   24 &    A & $   1.94 \pm    0.22$  \\
  AB327 & 1985-01-28 &       9 Sgr &   1733-130 & $   5.117 \pm    0.015$ & $ 13.4$ & $   17$ &   26 &    A & $   2.08 \pm    0.19$  \\
  AC116 & 1985-02-16 &       9 Sgr &   1733-130 & $   5.266 \pm    0.006$ & $ 13.4$ & $   19$ &   24 &    A & $   1.96 \pm    0.13$  \\
\multicolumn{4}{l}{{\em combined     AB327+AC116 }}            &                         &         &         &      &      & $   1.91 \pm    0.09$  \\
  AG178 & 1985-04-08 &       M8CO1 &   1751-253 & $  0.4826 \pm   0.0007$ & $  2.9$ & $   63$ &   25 &  BnA & $<        1.6$  \\
  AW158 & 1986-04-27 &     G5.97M8 &   1811-209 & $  0.3081 \pm   0.0006$ & $  3.9$ & $    4$ &   24 &    A & $   10.3 \pm     1.5$  \\
  AH259 & 1987-02-01 &          M8 &   1733-130 & $    6.24 \pm     0.02$ & $ 13.4$ & $  162$ &   24 &  DnC & $<         6.$  \\
  AH265 & 1987-04-24 &       M8CC1 &   1733-130 & $   6.580 \pm    0.017$ & $ 13.4$ & $   50$ &   26 &    D & $<         8.$  \\
   AT89 & 1988-01-22 &          M8 &   1733-130 & $   5.914 \pm    0.010$ & $ 13.4$ & $   62$ &   25 &    B & $    3.9 \pm    0.63$  \\
  AH265 & 1988-05-05 &       M8CC1 &   1751-253 & $   0.474 \pm    0.002$ & $  2.9$ & $   38$ &   23 &    C & $<         7.$  \\
  AB671 & 1993-01-21 &       9 Sgr &   1811-209 & $  0.3093 \pm   0.0009$ & $  3.8$ & $   19$ &   24 &    A & $   2.36 \pm    0.15$  \\
  AB671 & 1993-01-24 &       9 Sgr &   1811-209 & $  0.3142 \pm   0.0009$ & $  3.8$ & $   19$ &   26 &    A & $   2.34 \pm    0.28$  \\
  AB671 & 1993-01-29 &       9 Sgr &   1811-209 & $  0.3179 \pm   0.0004$ & $  3.8$ & $   19$ &   25 &  BnA & $   2.17 \pm    0.46$  \\
  AB671 & 1993-02-01 &       9 Sgr &   1811-209 & $  0.3140 \pm   0.0004$ & $  3.8$ & $   12$ &   25 &  BnA & $   2.13 \pm    0.24$  \\
  AB671 & 1993-02-14 &       9 Sgr &   1811-209 & $  0.3159 \pm   0.0005$ & $  3.8$ & $   10$ &   25 &  BnA & $<        20.$  \\
\multicolumn{4}{l}{{\em combined           AB671 }}            &                         &         &         &      &      & $   2.19 \pm    0.12$  \\
  AR328 & 1995-04-27 &       9 Sgr &   1733-130 & $   5.049 \pm    0.015$ & $ 13.4$ & $   11$ &   15 &    D & $<        30.$  \\
  AH557 & 1995-09-26 &          M8 &   1733-130 & $   5.470 \pm    0.005$ & $ 13.4$ & $   33$ &   26 &  BnA & $    3.4 \pm     0.9$  \\
  AW478 & 1997-10-03 &       9 Sgr &   1751-253 & $  0.4773 \pm   0.0008$ & $  2.9$ & $   13$ &   26 &  DnC & $<         9.$  \\
  AW478 & 1997-10-07 &       9 Sgr &   1751-253 & $  0.4778 \pm   0.0011$ & $  2.9$ & $   13$ &   26 &  DnC & $<         7.$  \\
  AW478 & 1997-10-11 &       9 Sgr &   1751-253 & $  0.4792 \pm   0.0006$ & $  2.9$ & $   13$ &   26 &  DnC & $<         6.$  \\
  AW478 & 1997-10-15 &       9 Sgr &   1751-253 & $  0.4773 \pm   0.0005$ & $  2.9$ & $   13$ &   26 &  DnC & $<         6.$  \\
  AW478 & 1997-10-18 &       9 Sgr &   1751-253 & $  0.4766 \pm   0.0014$ & $  2.9$ & $    8$ &   24 &  DnC & $<         6.$  \\
\multicolumn{4}{l}{{\em combined           AW478 }}            &                         &         &         &      &      & $<         5.$  \\
  AW515 & 1999-06-08 &       9 Sgr &   1820-254 & $  0.9992 \pm   0.0016$ & $  4.0$ & $   10$ &   25 &   AD & $<         3.$  \\
 AB1005 & 2001-03-08 &       9 Sgr &   1751-253 & $  0.4495 \pm   0.0009$ & $  2.9$ & $    8$ &   25 &    B & $    2.9 \pm    0.74$  \\
 AB1094 & 2003-07-05 &       18006 &   1820-254 & $   0.695 \pm    0.002$ & $  4.1$ & $   20$ &   26 &    A & $(    7.1 \pm     3.7)$  \\
\hline
\\\multicolumn{11}{c}{{\bf 20 cm}} \\
\hline
   BIGN & 1982-05-26 &       9 Sgr &   1733-130 & $    5.38 \pm     0.02$ & $ 13.4$ & $   72$ &   25 &    A & $   5.23 \pm    0.20$  \\
   AC42 & 1983-08-22 &       9 Sgr &   1733-130 & $   4.525 \pm    0.015$ & $ 13.4$ & $   31$ &   25 &    A & $   3.24 \pm    0.37$  \\
  AG163 & 1984-12-07 &         022 &   1751-253 & $   1.069 \pm    0.008$ & $  3.0$ & $    1$ &   22 &    A & $<        25.$  \\
  AC116 & 1984-12-21 &       9 Sgr &   1733-130 & $    5.86 \pm     0.02$ & $ 13.4$ & $   10$ &   26 &    A & $<        25.$  \\
  AR120 & 1985-01-24 &        G6.0 &   1811-209 & $   0.893 \pm    0.010$ & $  3.9$ & $   13$ &   25 &    A & $<        60.$  \\
  AR120 & 1985-01-26 &        G6.0 &   1811-209 & $   0.879 \pm    0.004$ & $  3.9$ & $    9$ &   26 &    A & $<        50.$  \\
\multicolumn{4}{l}{{\em combined           AR120 }}            &                         &         &         &      &      & $<        35.$  \\
  AR283 & 1993-01-10 &      G006-5 &   1751-253 & $   0.980 \pm    0.005$ & $  3.1$ & $   12$ &   26 &    A & $<        50.$  \\
  AT143 & 1993-02-20 &    0060-010 &   1751-253 & $   1.041 \pm    0.008$ & $  2.8$ & $    4$ &   26 &  BnA & $<        30.$  \\
  AR359 & 1996-08-29 &         M8E &   1733-130 & $   5.367 \pm    0.011$ & $ 13.6$ & $    6$ &   26 &    D & $<        35.$  \\
   AY76 & 1996-11-23 &         M8E &   1751-253 & $   1.033 \pm    0.003$ & $  3.1$ & $    4$ &   26 &    A & $<        25.$  \\
 AB1005 & 2001-03-08 &       9 Sgr &   1751-253 & $   1.177 \pm    0.002$ & $  2.9$ & $    8$ &   25 &    B & $   12.7 \pm     3.6$  \\
\hline
\\\multicolumn{11}{c}{{\bf 90 cm}} \\
\hline
  AJ194 & 1990-08-01 &     NGC6544 &   1830-360 & $   29.80 \pm     0.16$ & $ 12.2$ & $   90$ &   25 &    B & $<        50.$  \\
\hline
\end{longtable}
\tablefoot{
Column (1) gives the programme name, (2) the date of the observation,
(3) the source on which the observation was centred,
(4) the phase calibrator name (J2000 coordinates),
(5) the phase calibrator flux and 
(6) the distance of the phase calibrator to the observed target,
(7) the integration time on the source,
(8) the number of antennas that gave a usable signal,
(9) the configuration the VLA was in at the time of the observation, and
(10) the measured flux.
Upper limits are 3 $\times$ the RMS,
values between brackets indicate marginal detections (signal-to-noise 
ratio~$<3$). \\
Abbreviations used in column (3) 
are: [1] = J1803-244; [2] = 18039-24216.
}

\end{document}